\documentclass[12pt,preprint]{aastex}			
\slugcomment{To appear in the ApJL EVLA special Issue.}
\begin{document}



\title{High-resolution EVLA image of dimethyl ether (CH$_{3}$)$_{2}$O in Orion--KL}

	\author{C. Favre}
	\affil{Department of Physics and Astronomy, University of \AA rhus, Ny Munkegade 120, DK-8000 \AA rhus C, Denmark}
	\affil{Universit\'e de Bordeaux, OASU,  2 rue de l'Observatoire, BP 89, 33271 Floirac Cedex, France}
	\affil{ CNRS, UMR 5804, Laboratoire dÕAstrophysique de Bordeaux, 2 rue de lÕObservatoire, BP 89, 33271 Floirac Cedex, France}
	\email{favre@phys.au.dk}

	\author{H.~A. Wootten\altaffilmark{1} and A.~J.  Remijan\altaffilmark{1} }
	\affil{National Radio Astronomy Observatory, 520 Edgemont Road, Charlottesville, VA 22903-2475, USA}
	\email{awootten, aremijan@nrao.edu}

	\author{N. Brouillet}
	\affil{Universit\'e de Bordeaux, OASU,  2 rue de l'Observatoire, BP 89, 33271 Floirac Cedex, France}
	\affil{ CNRS, UMR 5804, Laboratoire dÕAstrophysique de Bordeaux, 2 rue de lÕObservatoire, BP 89, 33271 Floirac Cedex, France}
	\email{brouillet@obs.u-bordeaux1.fr}

	\author{T.~L. Wilson}
	\affil{Naval Research Laboratory, Code 7210, Washington, DC 20375, USA}
	\email{tom.wilson@nrl.navy.mil}

	\and

	\author{D. Despois and A. Baudry}
	\affil{Universit\'e de Bordeaux, OASU,  2 rue de l'Observatoire, BP 89, 33271 Floirac Cedex, France}
	\affil{ CNRS, UMR 5804, Laboratoire dÕAstrophysique de Bordeaux, 2 rue de lÕObservatoire, BP 89, 33271 Floirac Cedex, France}
	\email{despois, baudry@obs.u-bordeaux1.fr}

\altaffiltext{1}{The National Radio Astronomy Observatory is a facility of the National Science Foundation operated under cooperative agreement by Associated Universities, Inc.}



\begin{abstract}
We report the first sub-arc second (0.65$\arcsec$ $\times$ 0.51$\arcsec$) image of the dimethyl ether molecule, (CH$_{3}$)$_{2}$O, toward the Orion Kleinmann-Low nebula (Orion--KL). The observations were carried at 43.4~GHz  with the Expanded Very Large Array (EVLA). The distribution of the lower energy transition 6$_{1,5} - 6_{0,6}$, EE (E$\rm_{u}$ = 21~K) mapped in this study is in excellent agreement with the published dimethyl ether emission maps imaged with a lower resolution. The main emission peaks are observed toward the Compact Ridge and Hot Core southwest components, at the northern parts of the Compact Ridge and in an intermediate position between the Compact Ridge and the Hot Core. 
A notable result is that the distribution of dimethyl ether is very similar to that of another important larger O-bearing species, the methyl formate (HCOOCH$_{3}$), imaged at lower resolution. 
Our study shows that higher spectral resolution (WIDAR correlator) and increased spectral coverage provided by the EVLA  offer new possibilities for imaging complex molecular species. The sensitivity improvement and the other EVLA improvements make this instrument well suited for high sensitivity, high angular resolution, molecular line imaging.
\end{abstract}




\section{Introduction}
 \label{sec:Intro}

The Orion complex is among the most studied sources in our Galaxy. At a distance of $\sim$420~pc \citep[see, e.g.][]{Sandstrom:2007,Hirota:2007,Menten:2007} this is the nearest site of recent high mass star formation. 
The prominent \ion{H}{2} regions NGC~1976 or M42 and NGC~1977 are on the near side of a large molecular cloud associated with Lynds dark cloud L1640 \citep[see][]{Odell:2001}. At the rear of the \ion{H}{2} region is the Orion Molecular Cloud, OMC--1. The brightest molecular emission in OMC--1 is near the \ion{H}{2} region NGC~1976 \citep[see, e.g.][]{Tatematsu:1993}. \citet{Plume:2000} measured the J=5--4 transition of $\rm^{13}$CO over a 0.5$\degr$ by 2$\degr$ area with a 3.2$\arcmin$ resolution. The brightest emission has an extent of $\sim$15$\arcmin$ in the north-south by $\sim$5$\arcmin$ in the east-west direction. Within this region of extended emission is a prominent maximum, the Kleinmann-Low (KL) infrared nebula \citep[see, e.g.][]{Dougados:1993,Gezari:1998} which exhibits emission from complex molecules, masers, outflows and warm dust.  In addition to extended IR emission from warm dust, there are $\sim$20 compact near-IR sources found toward Orion--KL \citep{Dougados:1993}.

In quasi-thermal continuum emission from dust grains, Orion--KL shows an overall extent of 10$\arcsec$ in $\rm\alpha$ by 15$\arcsec$ in $\rm\delta$.
With higher angular resolution, the form has a \textquotedblleft V\textquotedblright \  shape, with the symmetry axis at position angle $\sim$20$\degr$ east of north. Within this region, there is fine structure, with molecular species peaking at different positions. This reflects abundance differences, not excitation effects. 

In the methyl formate (HCOOCH$_{3}$) image \citep[see Fig. 4 of][]{Favre:2011} and in the frequency-integrated maps of several molecular tracers  \citep[see Fig. 5 of][]{Guelin:2008} the molecular distribution is seen as a V--shape with each arm having an observed length of 12$\arcsec$ by 3$\arcsec$, with the major axis at a position angle $\sim$20$\degr$ east of north having an opening angle of 38$\degr$. From Plateau de Bure Interferometer \citep[PdBI,][]{Favre:2011} and EVLA data (A. Remijan, unpublished), the methyl formate emission taken with resolutions of 1.8$\arcsec$ $\times$ 0.8$\arcsec$ and $\sim$5$\arcsec$ peaks at $\alpha\rm_{J2000}$ = 05$\rm^{h}$35$\rm^{m}$14$\arcsec$09,  $\delta\rm_{J2000}$ = -05$\degr$22$\arcmin$36$\arcsec$7. This is the center position of the compact ridge region. The compact ridge shows a larger abundance of oxygen-containing species. In CH$\rm_{3}$CH$\rm_{2}$CN, ethyl cyanide, the center of the Hot Core peaks at $\alpha\rm_{J2000}$ = 05$\rm^{h}$35$\rm^{m}$14$\arcsec$6,  $\delta\rm_{J2000}$ = -05$\degr$22$\arcmin$29$\arcsec$, with observed sizes of 9$\arcsec$ by 4$\arcsec$ at  P.A. $\sim$20$\degr$ east of north. This region shows an excess of nitrogen bearing species \citep[see, e.g.][]{Blake:1996,Friedel:2008,Wilson:2000,Wright:1996}. The chemistry of Orion--KL is rich, with rather large abundances of complex molecules. There are two possibilities: (1) gas-phase reactions between ionized and neutral species, and (2) formation on dust grains, followed by liberation from grains by radiation from nearby IR sources. 

In addition to extended dust and molecular emission, Orion--KL also has a number of radio continuum sources \citep{Garay:1987,Churchwell:1987,Menten:1995,Rodriguez:2005}. Accurate high angular resolution radio data show that the highly obscured source \textquotedblleft I\textquotedblright , and the Becklin-Neugebauer object \citep[BN][]{Becklin:1967} show large proper motions \citep{Plambeck:1995}. 
Subsequently, \citep{Rodriguez:2005,Gomez:2005,Gomez:2008,Goddi:2011} have found proper motions of source I, BN and the radio counterpart of source Ò\textquotedblleft n\textquotedblright. 

Given the large amount of activity in the Orion--KL region, high angular resolution is needed to separate the various influences. There are two coupled problems to be solved. The first is the chemistry of the Hot Core and the Compact Ridge. Although the centers of these regions are separated by only $\sim$10$\arcsec$ (0.02~pc), their chemistry is quite different. In addition, models must accurately predict the abundances of complex species and their survival in the presence of the activity as shown by the large proper motions of sources BN, I and n. Given these questions, the first task is to image complex molecular species with arc second or better angular resolutions, to test chemistry models. For this purpose, we have imaged the KL nebula in dimethyl ether (CH$_{3}$)$_{2}$O, which had been mapped with lower spatial resolution by \citet{Friedel:2008} and \citet{Guelin:2008}.

In Sect. \ref{sec:Observations} we present our observations and we briefly describe the data reduction methods. Our maps are presented in Sect. \ref{sec:Results}. We compare our results to previous studies in Section \ref{sec:Comparison}. Conclusions are presented in the last Section.
 


\section{Observations}
 \label{sec:Observations}

A quartet of dimethyl ether ((CH$_{3}$)$_{2}$O) lines (see Table \ref{Tab.spectro-param}) were observed on 9 Nov 2010 (1 hour session) and 19 Jan 2011 (4 hours session) with the NRAO EVLA \citep[see][]{Perley:2011} in the \textquotedblleft C\textquotedblright \ configuration (\textit{proposal code: 10B-223}). Dual polarization observations were made toward a single pointing ($\alpha\rm_{J2000}$ = 05$\rm^{h}$35$\rm^{m}$14$\arcsec$2, $\delta\rm_{J2000}$ = -05$\degr$22$\arcmin$36$\arcsec$0), tuned to the frequency of 43433.5449~GHz.
Here we discuss the 19 Jan 2011 observations. The targeted lines were the 6$_{1,5} - 6_{0,6}$ lines lying near 43.44757~GHz. AE, EA, AA and EE lines appear close together in a distinctive \textquotedblleft pitchfork\textquotedblright \ pattern.  The upper energy levels are only 21~K above ground and easily excited. 

Winds were low and a few scattered clouds decorated the sky under dry conditions on 19 Jan. Using the measured weather parameters and an atmospheric model generated from the ATM program within CASA we estimate that the atmospheric opacity was 0.058 during the observations; we have corrected the data for that assumed value.

The full width at half power (FWHP) at the observing frequency of the primary beam of each antenna is $\sim$1$\arcmin$. The WIDAR spectrometer provided 256 channels over a 16~MHz bandwidth and a channel spacing of 62.5~kHz (0.43~km~s$^{-1}$) in each of the two orthogonal polarizations. Centered near the nominal velocity of the ambient cloud, v$\rm_{LSR}$=8~km~s$^{-1}$, 111.2~km~s$^{-1}$ of velocity coverage was available in total.

J0541-0541 (measured flux density S$_{43.43}$= 0.386$\pm$0.001 Jy) served as the gain calibrator for Orion--KL using fast-switching (2$\rm^{m}$ on source, 1$\rm^{m}$ on the calibrator). Data calibration followed standard high-frequency procedures in CASA, including using a model for the brightness distribution of the absolute flux calibrator J0137+331 (3C48), which gives a flux density of S$_{43.43}$= 0.528~Jy. 
J0319+4130 (3C84) was used as bandpass calibrator (S$_{43.43}$=13.6 $\pm$0.5~Jy). 

The continuum emission was subtracted from the January dataset and imaged.  
The final continuum and continuum-subtracted images for analysis have been corrected for the response of the primary beam.

The final naturally weighted maps reached a (5$\sigma$) sensitivity of 20~mJy per beam (39~K) in the line images and 3~mJy per beam in the continuum images, as expected. The beamsize is 0$\farcs$65 $\times$ 0$\farcs$51 at position angle 38$\degr$.



\section{Results}
\label{sec:Results}

\subsection{Continuum emission}

\begin{figure}[h!]
\epsscale{1.}
\plotone{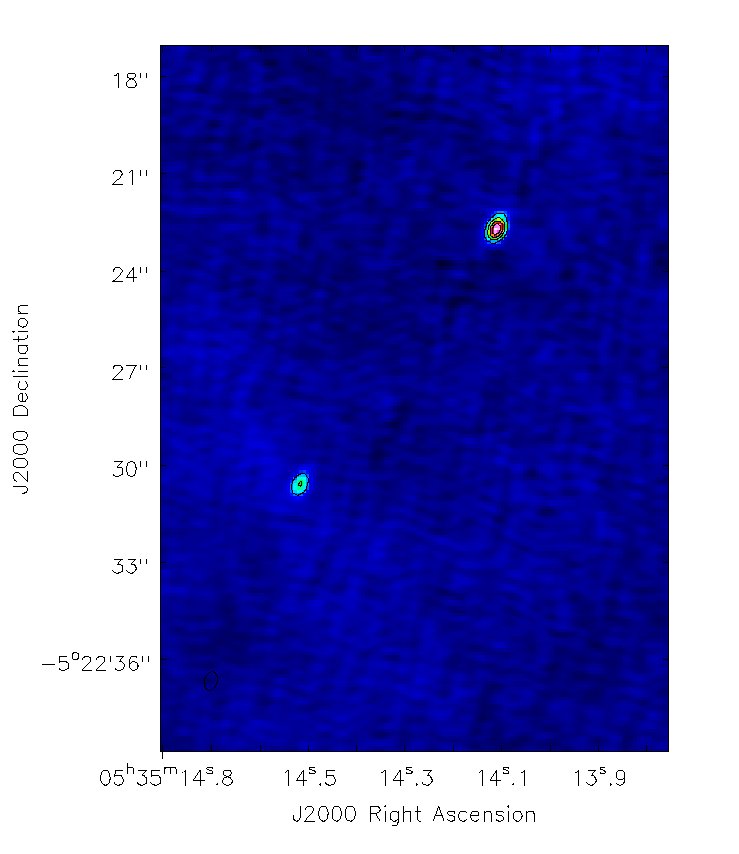}
\caption{43~GHz continuum emission observed with the EVLA toward Orion BN/KL. The first contour and the level step are 4.7~mJy~beam$^{-1}$ and the clean beam (0.65$\arcsec$ $\times$ 0.51$\arcsec$) is shown in the bottom left corner. Position of the BN object is ($\alpha\rm_{J2000}$ = 05$\rm^{h}$35$\rm^{m}$14$\fs$1094, $\delta\rm_{J2000}$ = -05$\degr$22$\arcmin$22$\farcs$724) and the position of the radio source I  is ($\alpha\rm_{J2000}$ = 05$\rm^{h}$35$\rm^{m}$14$\fs$5141, $\delta\rm_{J2000}$ = -05$\degr$22$\arcmin$30$\farcs$575) \citep{Goddi:2011}.}
\label{fig.continuum}
\end{figure}

The 43~GHz continuum emission observed toward Orion~BN/KL is shown in Fig. \ref{fig.continuum}. Two continuum sources, the BN object and the radio source I, are detected above 5$\sigma$ with a peak flux density per synthesized beam S$\rm_{\nu}$, of 22.3~mJy~beam$^{-1}$ and 9.8~mJy~beam$^{-1}$, respectively. Some of the structure of the Hot Core source is resolved. Emission from the BN object also appears quite compact. The total flux densities S$\rm_{\nu}$, are given in Table \ref{Tab.cont}.

\subsection{Dimethyl ether (CH$_{3}$)$_{2}$O} 

\begin{figure}[h!]
\vspace*{10cm} 
\includegraphics[angle=270,width=\textwidth,bb=557 87 40 754,clip=]{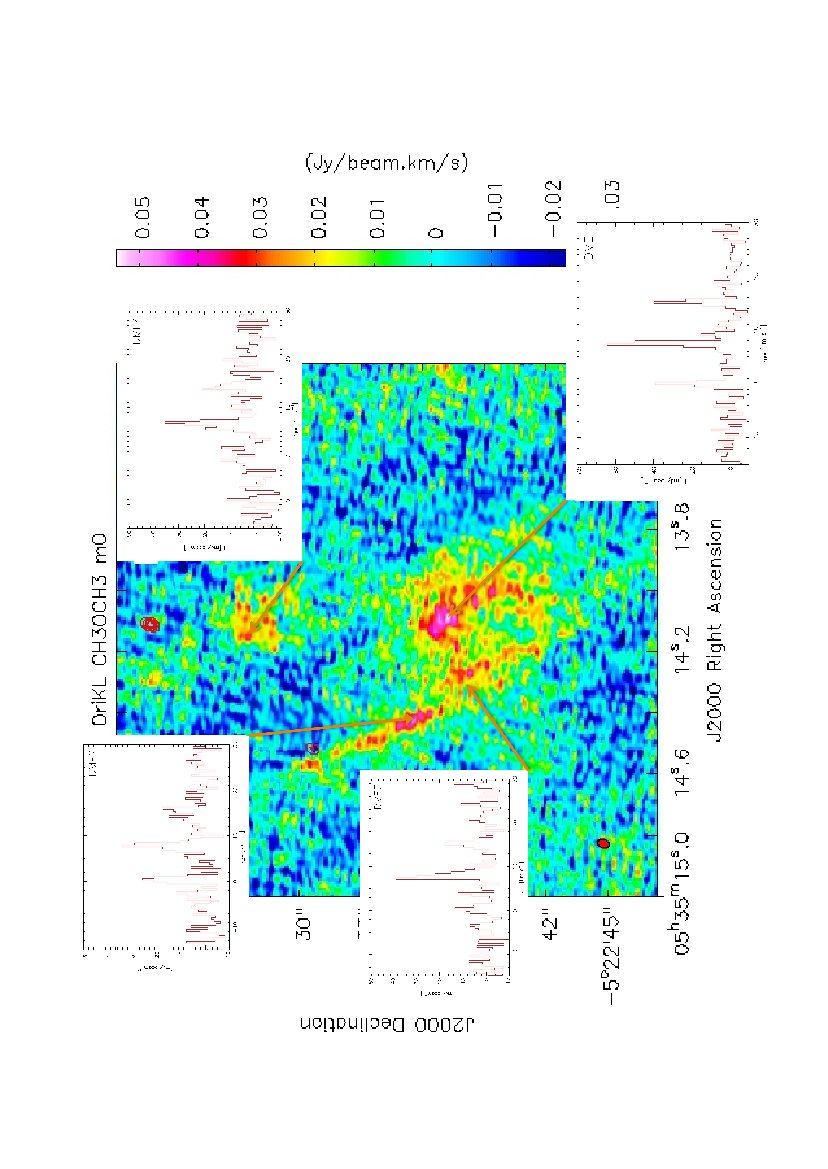}
\caption{Dimethyl ether (CH$_{3}$)$_{2}$O emission map observed with the EVLA toward Orion BN/KL. The synthesized beam, 0.65$\arcsec$ $\times$ 0.51$\arcsec$, is shown in red in the lower left corner. The red contours show the 43~GHz continuum emission of the BN object and the radio source I (see Fig. \ref{fig.continuum}). Spectra displayed in this figure are observed toward the four main (CH$_{3}$)$_{2}$O emission peaks (DME1 to DME4), i.e. in the following directions:  the Compact Ridge (bottom right), the Hot Core-SW (top left), the intermediate region linking the Hot Core to the Compact Ridge (bottom left) and the north of the Compact Ridge (top right).
}
\label{fig.mapdme}
\end{figure}

The dimethyl ether emission map at 43.447~GHz is shown in Fig. \ref{fig.mapdme}. 
The observed four main molecular emission peaks, hereafter DME1, DME2, DME3 and DME4, are located toward the Compact Ridge, the Hot Core southwest (Hot Core-SW), in an intermediate position between the Compact Ridge and the Hot Core-SW and in the north of the Compact Ridge, respectively. Most of the peaks appear with a LSR velocity of 7.4~km~s$^{-1}$. As for many O-bearing species the emission distribution presents an extended V-shape linking the radio source I to the BN object \citep[see, e.g.][]{Guelin:2008}. 

The high spectral resolution provided by the WIDAR correlator allows us to distinguish between the AA and EE transitions, although the AE and EA transitions are blended (see Table \ref{Tab.spectro-param}).
Spectra of the lower energy transitions 6$_{1,5} - 6_{0,6}$,~AA, EE and AE/EA~(E$\rm_{u}$ = 21~K) observed toward the regions DME1 to DME4 are presented in Fig. \ref{fig.mapdme}.
The observed line parameters (velocity, intensity) of the detected (CH$_{3}$)$_{2}$O~6$_{1,5} - 6_{0,6}$,~EE transition are summarized in Table \ref{Tab.dme-param}.



\section{Comparison with other results and analysis }
 \label{sec:Comparison}

\subsection{Comparison with continuum maps}

The total flux densities S$\rm_{\nu}$ (see Table \ref{Tab.cont}), obtained toward source I, in the Hot Core region, and the BN object are in very good agreement with the 43~GHz Very Large Array (VLA) values reported by \citet{Menten:1995,Chandler:1997,Reid:2007,Goddi:2011}. 

\subsection{Single dish observations}

Green Bank Telescope (GBT) molecular surveys have been undertaken from 42.3~GHz to 43.6~GHz \citep{Goddi:2009a} and from 42.7~GHz to 45.6~GHz (Wootten, unpublished and Gu\'elin et al. 2008). A comparison between these surveys and our EVLA spectra has allowed us to conclude to within the uncertainties, that all flux measured in either GBT survey for the (CH$_{3}$)$_{2}$O, 6$_{1,5} - 6_{0,6}$ transition is recovered in our EVLA images.

The excitation of the dimethyl ether emission may be estimated from fitting a rotational temperature to the three lines spanning an energy range 21-147~K from \citet{Goddi:2009a} or from two lines spanning the same range from \citet{Guelin:2008}.  
From these data, line sets observed simultaneously on the GBT, we estimate T$\rm_{rot}$((CH$_{3}$)$\rm_{2}$O)$\sim$120~K, in agreement with estimates from transitions of HCOOCH$_{3}$ of \citet{Favre:2011}.
Note that the brightness temperature of the dimethyl ether line reported here toward the peak of the compact ridge (DME1, see Fig.~\ref{fig.mapdme}) is $\sim$140~K and that this line is optically thick ($\tau$ estimated to be 2.9), consistent with the rotational temperature in the broader (16$\arcsec$) GBT beam. Using lines of acetone (CH$_{3}$)$_{2}$CO in the same GBT spectrum, \citet{Goddi:2009a} found a rotational temperature twice this value, suggesting an origin in the Hot Core.  Higher resolution images are needed to understand how the distributions of otherwise similar O-bearing molecules differ, and to glean lessons on how large molecules may form and vanish in molecular clouds.

\subsection{Interferometric observations}

Using the Combined Array for Research in Milimeter-Wave Astronomy (CARMA), \citet{Friedel:2008} have imaged the 
dimethyl ether transition 13$_{0,13} - 12_{1,12}$ (E$\rm_{u}$ = 86~K) with a beam size of 2.5$\arcsec$ $\times$ 0.85$\arcsec$. Our observed (CH$_{3}$)$_{2}$O emission peaks (DME1, DME2 and DME4, see Fig.~\ref{fig.mapdme}) are present in their lower resolution data \footnote{Note that the Compact Ridge position defined in \citet{Friedel:2008} is not the the same as that used in this study \citep[e.g. reference taken from][]{Beuther:2005}. Their IRc5 position is closer to our Compact Ridge position (1.5$\arcsec$ away).}. 
CARMA observations reveal a similar LSR velocity at 7.6~km~s$^{-1}$. However, their lower spectral resolution (by a factor of 1.4 compared to that of ours) did not allow them to resolve the AA, EE and AE/EA lines.

A comparison between our EVLA map and the Plateau de Bure Interferometer (PdBI)  dimethyl ether transition 12$_{2,10} - 11_{3,9}$ (E$\rm_{u}$ = 78~K) imaged by \citet{Guelin:2008}\footnote{Note that a misprint with the dimethyl ether quantum numbers appears in the Table 1 of \citet{Guelin:2008}.}, with a synthesized beam of 3.3$\arcsec$ $\times$ 1.7$\arcsec$, shows very good agreement. With a higher-angular resolution the distribution emission of the (CH$_{3}$)$_{2}$O  mapped with EVLA presents the same main strongest emission peaks within the entire region: DME1, DME2 and DME4 \citep[regions A, B and C, respectively in Fig. 5 of][]{Guelin:2008}.

Our EVLA sub-arc second (0.65$\arcsec$ $\times$ 0.51$\arcsec$) image of the dimethyl ether clearly reveals a new emission peak DME3, located between the Hot Core-SW and the Compact Ridge (see Fig. \ref{fig.mapdme}). 

\subsection{Relation to methyl formate (HCOOCH$_{3}$) and shocks}

There is a debate whether dimethyl ether is formed on grain surfaces or in the gas phase via a path involving methanol as a precursor \citep{Garrod:2008,Peeters:2006}. This largest O-bearing molecule is highly abundant in star formation region, particularly in hot cores
\citep{Nummelin:2000,Sutton:1995,Turner:1991}. Studies of the astrochemistry of this molecule, as well as its spatial distribution will bring important data for understanding high-mass star forming regions.

A comparison between the dimethyl ether distribution with that of another important O-bearing molecule, methyl formate HCOOCH$_{3}$  \citep[see, e.g.][]{Favre:2011} shows that both of these species peak at the same locations within the same velocity range. In particular, these share common emission peaks in the north of the Compact Ridge, in the Compact Ridge itself, toward the Hot Core-SW and toward an intermediate region linking the Hot Core to the Compact Ridge. These four emission regions are shown in Fig. \ref{fig.mapdme} \citep[see labels DME1 to DME4 in the present letter and MF1 to MF4/5 in][]{Favre:2011}. 
Hence the formation of these two molecules must have some relation, regardless whether the mechanism involves gas phase or grain surface formation.

\citet{Favre:2011} have shown that shocks could be responsible for the HCOOCH$_{3}$ production in the Compact Ridge region. In particular, the clear association observed  between the emissions of the 2.12~$\micron$ vibrationally excited H$\rm_{2}$ \citep{Lacombe:2004} and the methyl formate could result from shocks between the interstellar material and bullets owing to the close encounter of the sources I, n and BN 500~years ago \citep{Zapata:2009,Bally:2011}.
The similar spatial distribution observed  suggests that the same mechanisms could be at the origin of the release of (CH$_{3}$)$_{2}$O (itself or a precursor) from ice grain mantles.



\section{Conclusion }
 \label{sec:Conclusion}
 
We have imaged the distribution of the dimethyl ether 6$_{1,5} - 6_{0,6}$, EE (E$\rm_{u}$ = 21~K) transition with the high-angular resolution of 0.65$\arcsec$ $\times$ 0.51$\arcsec$ using the EVLA. The high resolution provided by the WIDAR correlator, allowed us to spectrally separate transitions AA, EE and AE/EA.
Our study shows the great capabilities offered by this new interferometer at such high spatial and spectral resolution.

Our results confirm the three main emission peaks of (CH$_{3}$)$_{2}$O observed in the previous studies with a lower angular resolution. A new dimethyl ether emission peak has been identified thanks to the high spatial resolution of these EVLA observations. The similarity between spatial distributions of this molecule and HCOOCH$_{3}$ and vibrationally excited H$_{2}$ toward the Compact Ridge suggest that same production mechanisms that is, shocks may be involved.



\acknowledgments
We are grateful to the entire EVLA staff who produced such an excellent instrument. 

{\it Facilities:} \facility{The Expanded Very Large Array of the National Radio Astronomy Observatory.}



\bibliographystyle{apj}
\bibliography{biblio.bib}




\clearpage

\begin{table}[h!]
\begin{center}
\caption{Spectroscopic line parameters\tablenotemark{a} of dimethyl ether (CH$_{3}$)$_{2}$O.}
\label{Tab.spectro-param}
\begin{tabular}{ccccc}
\tableline\tableline
 Frequency\tablenotemark{b} &Transition & $\langle$ S$\rm_{i,j}$$\mu$$^{2}$ $\rangle$& E$\rm_{u}$ \\
  (MHz) && (D$^{2}$)& (K)   \\
 \tableline
43446.4708(12)& 6$_{1,5} - 6_{0,6}$~AE & 14.0 & 21.1\\
43446.4713(12)& 6$_{1,5} - 6_{0,6}$~EA & 28.1 & 21.1\\
43447.5415(11)& 6$_{1,5} - 6_{0,6}$~EE & 112.2 & 21.1 \\
43448.6120(14) & 6$_{1,5} - 6_{0,6}$~AA & 42.1 & 21.1 \\
\tableline    \tableline
\end{tabular}
\tablenotetext{a}{All spectroscopic data from (CH$_{3}$)$_{2}$O taken from \citet{Groner:1998} available from the JPL molecular line catalog \citep{Pickett:1998} at Splatalogue \citep[www.splatalogue.net,][]{Remijan:2007}.}
\tablenotetext{b}{Errors are 2$\sigma$.}
\end{center}
\end{table}

\begin{table}[h!]
\begin{center}
\caption{Total flux densities obtained at 43~GHz for the sources BN and I.}
\label{Tab.cont}
\begin{tabular}{ccccc}
\tableline\tableline
Source & $\alpha\rm_{J2000}$\tablenotemark{a}& $\delta\rm_{J2000}$\tablenotemark{a}& Uncertainty\tablenotemark{b} &S$\rm_{\nu}$ \\
 & ($\rm^{h}$ $\rm^{m}$ $\fs$)& ($\degr$ $\arcmin$ $\farcs$)& (mJy) & (mJy) \\
 \tableline
BN & 05~35~14.1094& -05~22~22.724 & 3 & 29\\
I & 05~35~14.5141& -05~22~30.575 & 3 & 16 \\
\tableline
\end{tabular}
\tablenotetext{a}{\citet{Goddi:2011}.}
\tablenotetext{b}{5$\sigma$.}
\end{center}
\end{table}

\begin{table}[h!]
\begin{center}
\caption{Line parameters for the detected (CH$_{3}$)$_{2}$O 6$_{1,5} - 6_{0,6}$,~EE transition at 43.4~GHz toward Orion BN/KL.}
\label{Tab.dme-param}
\begin{tabular}{llccc}
\tableline\tableline
Region & Peak label & v$\rm_{LSR}$ & $\Delta$v$\rm_{LSR}$ &I$\rm_{0}$ \\
 & &  (~km~s$^{-1}$)  &  (~km~s$^{-1}$) &(Jy~beam$^{-1}$) \\
 \tableline
Compact Ridge (CR)& DME1&7.4 &0.9 & 64\\
Hot Core-SW (HC-SW)&DME2&7.4 &1.9 &14 \\
Intermediate of CR and HC-SW& DME3 &  7.4 &1.7 &35 \\
Compact Ridge North &DME4& 6.9& 1.5& 40\\
\tableline    \tableline
\end{tabular}
\end{center}
\end{table}


\end{document}